# EFFECTS OF MISALIGNMENTS IN THE OPTICAL VORTEX TRANSFORMATION PERFORMED BY HOLOGRAMS WITH EMBEDDED PHASE SINGULARITY

A.Ya. Bekshaev*, S.V. Sviridova

*I.I. Mechnikov National University, Dvorianska 2, 65082, Odessa, Ukraine*

**Abstract**

Spatial characteristics of diffracted beams produced by a "fork" hologram from an incident circular Laguerre-Gaussian beam whose axis differ from the hologram optical axis are studied theoretically. General analytical representations for the complex amplitude distribution of a diffracted beam are derived in terms of superposition of Kummer beams or hypergeometric-Gaussian beams. The diffracted beam structure is determined by combination of the "proper" topological charge $m$ of the incident vortex beam and the topological charge $l$ of the singularity "imparted" by the hologram. Evolution of the diffracted beam structure is studied in detail for several combinations of $m$ and $l$ and for various incident beam displacements with respect to the optical axis of the hologram. Variations of the intensity and phase distribution due to the incident beam misalignment are investigated and possible applications for the purposeful optical-vortex beam generation and optical measurements are discussed.

**1. Introduction**

During past decades, light beams with optical vortices (OV) have become popular objects of the physical optics [1–6]. Their peculiar physical properties (wavefront singularities, isolated points and/or compact regions of zero intensity, circulation energy fluxes, etc.) are coupled with promising applications. In particular, they concern capturing and manipulation of microparticles [7–9], measurement of small displacements and improved resolution of neighboring astronomical objects 2[10–15], encoding and processing information [16–24] (including analyses of quantum entanglement and verification of the Bell inequalities [19–24]). Both generation and analysis of the OV beams can be suitably performed with assistance of computer-generated holograms (CGH) whose special feature is the groove bifurcation ("fork" structure, see Fig. 1) [16,19,24,25–34].

---

* Corresponding author. Tel.: +38 048 723 80 75
  *E-mail address*: bekshaev@onu.edu.ua (A.Ya. Bekshaev)

Properties of the OV beams produced with the help of such CGHs were studied in detail in a series of works. Most results relate to the common case when the incident beam is vortexless (Gaussian) [28–39]; as to transformations of beams already possessing OV, important for the OV analysis [19–25], the systematic study is now at the initial stage [40]. If an OV beam with topological charge $m$ falls onto a CGH where single groove divides into $|q| + 1$ branches (in Fig. 1 $q = 1$), the $n$-order diffracted beam acquires the OV with topological charge

$$l = m + qn. \tag{1}$$

The signed integer number $q$ is usually referred to as the topological charge of the phase singularity "embedded" in the CGH [34–36]; the product $qn$ is thus the topological charge of the OV "imparted" to the beam by the hologram.

However, this simple picture is only correct for the 'nominal' situations when the incident beam axis is normal to the hologram plane and crosses it in the bifurcation point (center of the CGH) [40]. In many applications this condition cannot be realized exactly. Every practical configuration contains inevitable deviations from the perfect scheme, and knowledge of the role and significance of these deviations is highly desirable for the design and development of optical systems with CGH. Additionally, this knowledge can be used for deliberate control of characteristics of OV beams formed with the help of CGH. Presumable metrological applications [13–15] and employment of special versions of CGH for discriminating different OV beams in systems of information transfer and processing [16–19,22,23] also crucially depend on the misalignment effects.

In this work, we continue to analyze transformations of the OV beams in CGHs with the "fork" structure following to the previously developed scheme [35,36,39,40], but with explicit allowance for possible geometrical misalignment between the incident beam and the CGH (see Fig. 1). The CGH is considered as a planar transparency with spatially inhomogeneous transmittance $T(\mathbf{r}_a)$ where $\mathbf{r}_a = (x_a, y_a) = (r\cos\phi, r\sin\phi)$ is the transverse radius-vector; the coordinate frame is chosen so that its origin coincides with the bifurcation point and axis $y_a$ is parallel to the grating grooves far from the "fork". The nominal axis of a readout (incident) beam coincides with axis $z_a$ forming a 3D Cartesian frame with axes $x_a$ and $y_a$ (at the grating plane $z_a = 0$).

In these coordinates, the transmittance function can be represented by the Fourier series,

$$T(\mathbf{r}_a) = \sum_{n=-\infty}^{\infty} T_n \exp\left[in\left(\frac{2\pi}{d}x_a + q\phi\right)\right] \tag{2}$$

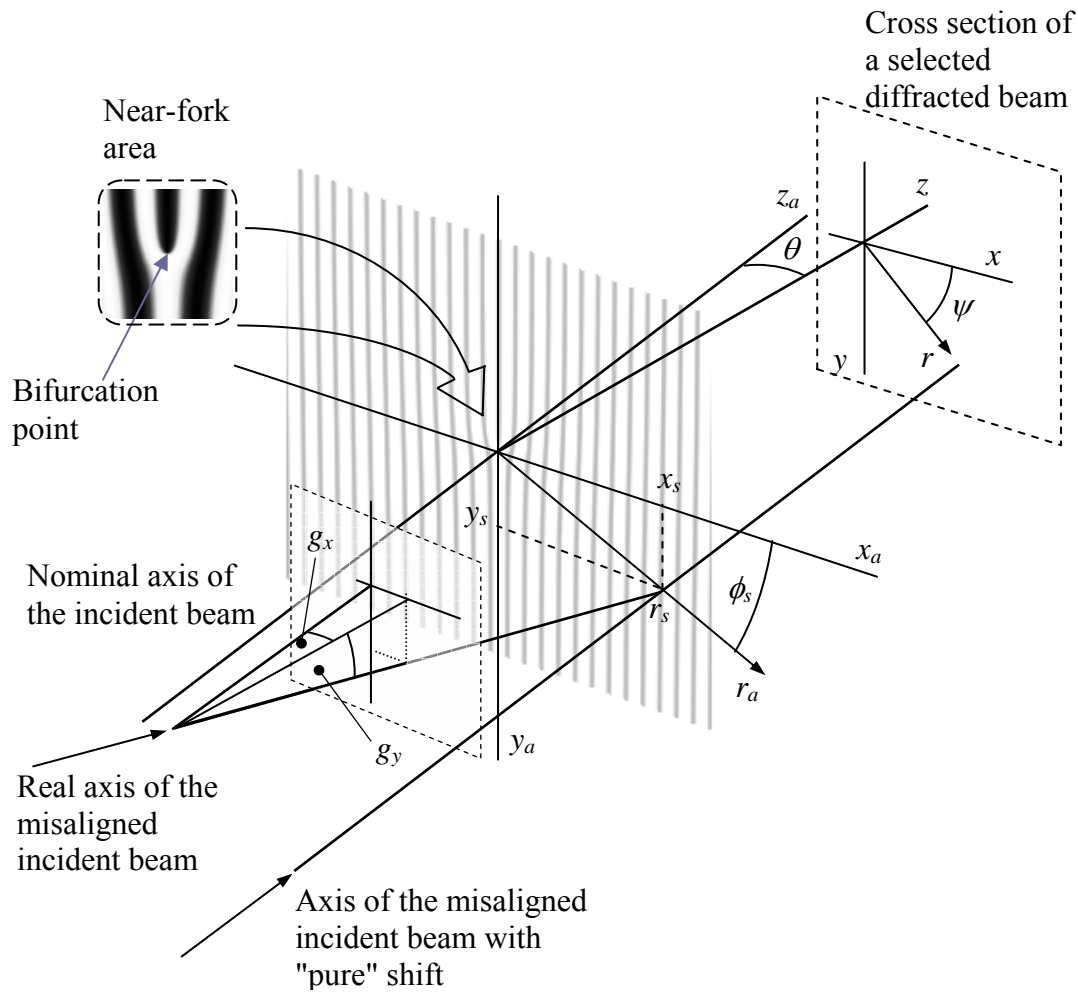

Fig. 1. Geometrical conditions of the beam transformation in the CGH. Nominal axis of the incident beam is axis $z_a$, intersecting the grating plane in the bifurcation point as is shown in the inset. Real axis of the incident beam is inclined and shifted; the diffracted beam nominal axis $z_a$ is deflected by angle $\theta$. Further explanations see in the text.

where $d$ is the grating period. Let the incident beam be monochromatic with the wave number $k$; then behind the grating, in accordance with expansion (2), a set of diffracted light beams (diffraction orders) is formed. In the nominal conditions, diffracted beams propagate in directions specified by condition [41]

$$\sin\theta = \frac{2\pi n}{kd}.  \qquad (3)$$

In fact, $n$-th term of expansion (2) produces the $n$-th diffraction order. For description of the field of a separate diffraction order, the associated coordinate frame $(x, y, z)$ is introduced, origin being in the bifurcation point and axis $z$ coinciding with the nominal axis of the diffracted beam (in Fig. 1, axes *x* and *y* are translated along axis *z* to the current cross section). Corresponding polar coordinates are determined as usual,

$$r = \sqrt{x^2 + y^2}, \quad \psi = \arctan\left(\frac{y}{x}\right).$$

Besides, in agreement with the usual practice, we assume that actual diffraction angles $\theta$ are small enough so that $\cos\theta \approx 1$ (see Fig. 1). This requirement implies that the considered CGH has low spatial frequency (practically, below 100 grooves per millimeter), and only low diffraction orders are admissible.

## 2. Mathematical formulation and analysis

As was shown before [35,36,40], a diffracted beam of a separate order propagates along axis *z* as a paraxial beam and its field can therefore be represented as $E(x,y,z) = u(x,y,z)\exp(ikz)$ with the slowly varying complex amplitude $u(x,y,z)$ [42]. Evolution of the diffracted beam can be suitably described in dimensionless variables where the transverse coordinates $\mathbf{r}_a = (x_a, y_a)$ and $\mathbf{r} = (x, y)$ are measured in units of the characteristic transverse size of the beam *b* while $z_a$ and *z* are measured in units of the associated longitudinal scale of the beam $kb^2$ [40]. Then, under the accepted small-angle diffraction conditions, the complex amplitude of the *n*-th diffraction order is determined by equation [35,40]

$$u_l(r,\psi,z) = \frac{1}{2\pi i z} \int u_a(r_a,\phi)\exp(il\phi)\exp\left\{\frac{i}{2z}\left[r_a^2 + r^2 - 2r_a r\cos(\phi-\psi)\right]\right\} r_a\, dr_a\, d\phi \qquad (4)$$

where *l* is defined via Eq. (1) and $u_a(r_a, \phi)$ is the complex amplitude distribution of the incident beam at $z_a = 0$. Eq. (4) determines $u_l(r,\psi,z)$ within a constant factor responsible for the transformation efficiency; in the present form, it is supposed to equal the unity, as if the whole input beam energy is transmitted to the chosen output beam.

In the previously considered case of ideal alignment (nominal geometric conditions of transformation) [40], the diffracted beam complex amplitude was calculated from Eq. (4) for the incident Laguerre-Gaussian beam $LG_{0m}$ in the form

$$u_a(x_a, y_a) \equiv \frac{1}{\sqrt{|m|!}}\left[x_a + i\,\mathrm{sgn}(m)\, y_a\right]^{|m|}\exp\left(-\frac{x_a^2 + y_a^2}{2}\right) = \frac{r_a^{|m|}}{\sqrt{|m|!}}\exp\left(-\frac{r_a^2}{2} + im\phi\right). \qquad (5)$$

Here the transverse unit of length *b* equals to the Gaussian envelope radius at the intensity level $e^{-1}$ of maximum and the longitudinal unit of length is the corresponding Rayleigh range; multiplier $\left(|m|!\right)^{-1/2}$ is added for convenience and provides the beams of every *m* with the same total power

$$\int_0^\infty |u_{lm}(r_a,\phi,z_a)|^2 r_a dr_a \int_0^{2\pi} d\phi = \pi. \qquad (6)$$

Now we consider situations when the incident beam is slightly displaced from the nominal position: namely, its axis crosses the grating plane in point $x_a = x_s$, $y_a = y_s$ and is inclined by small angles $(g_x, g_y)$ (see Fig. 1). As was shown before [36], all the effects of the beam inclination can be expressed by additional phase factors and by corresponding parallel translation of the complex amplitude distribution calculated in assumption $g_x = g_y = 0$. Consequently, we can restrict our consideration to the case of 'pure' transverse shift of the incident beam when

$$u(x_a, y_a) \to u_a(x_a - x_s, y_a - y_s)$$

so that

$$u_a(x_a, y_a) = \frac{1}{\sqrt{|m|!}}\left[(x_a - x_s) + i\,\mathrm{sgn}(m)(y_a - y_s)\right]^{|m|} \exp\left(-\frac{|\mathbf{r}_a - \mathbf{r}_s|^2}{2}\right) \qquad (7)$$

where

$$|\mathbf{r}_a - \mathbf{r}_s|^2 = (x_a - x_s)^2 + (y_a - y_s)^2 = r_a^2 + r_s^2 - 2r_a r_s \cos(\phi - \phi_s) \qquad (8)$$

is the squared absolute value of the incident beam displacement, $\phi_s$ specifies the displacement direction. Further, the incident beam complex amplitude distribution (7) can be reduced to the form

$$u_a(x_a, y_a) = u_a(r_a,\phi) = \frac{1}{\sqrt{|m|!}}\left(r_a e^{i\,\mathrm{sgn}(m)\phi} - r_s e^{i\,\mathrm{sgn}(m)\phi_s}\right)^{|m|} \exp\left(-\frac{|\mathbf{r}_a - \mathbf{r}_s|^2}{2}\right)$$

$$= \frac{1}{\sqrt{|m|!}} \sum_{\mu=0}^{|m|} \frac{|m|!}{\mu!(|m|-\mu)!}\left[-r_s e^{i\phi_s \mathrm{sgn}(m)}\right]^{|m|-\mu}\left[r_a e^{i\phi\,\mathrm{sgn}(m)}\right]^{\mu} \exp\left[-\frac{r_a^2 + r_s^2}{2} + r_a r_s \cos(\phi - \phi_s)\right]. \qquad (9)$$

This expression enables us to perform analytical investigation of the diffracted beam propagation in the way similar to the approach used for the perfectly aligned configuration [40]. Substitution of (9) into integral (4) directly gives

$$u_l(r,\psi,z) = \frac{1}{z}\exp\left(\frac{i}{2z}r^2 - \frac{1}{2}r_s^2\right)\sum_{\mu=0}^{|m|}\frac{\sqrt{|m|!}}{\mu!(|m|-\mu)!}\left[-r_s e^{i\phi_s \mathrm{sgn}(m)}\right]^{|m|-\mu} u_{l\mu}(r,\psi,z) \qquad (10)$$

where

$$u_{l\mu}(r,\psi,z) = \frac{1}{2\pi i}\int_0^\infty r_a^{\mu+1}\exp\left[-\frac{r_a^2}{2}\left(1 - \frac{i}{z}\right)\right]dr_a \int_{-\pi}^{\pi}\exp\left[i\nu\phi - iQ\frac{r_a}{z}\cos(\phi - \gamma)\right]d\phi, \qquad (11)$$

$$\nu = l + \mu\,\mathrm{sgn}(m), \qquad (12)$$

$$Q\cos\gamma = r\cos\psi + izr_s\cos\phi_s = x + izx_s, \quad Q\sin\gamma = r\sin\psi + izr_s\sin\phi_s = y + izy_s. \qquad (13)$$

Integration over the polar angle in Eq. (11) is easily performed due to known relation for Bessel functions $J_\nu$ with integer index $\nu$ [43]

$$\int_{-\pi}^{\pi} \exp\left[i\nu\phi - iQ\frac{r_a}{z}\cos(\phi-\gamma)\right]d\phi = 2\pi(-i)^{|\nu|} e^{i\nu\gamma} J_{|\nu|}\left(Q\frac{r_a}{z}\right).$$

Then integral (11) can be found with the help of the Tables of integrals (see [44], pt. 2.12.9), and by using the Kummer transformation [43,46] one finally obtains

$$u_{l\mu}(r,\psi,z) = C_{l\mu}\left(\frac{z}{z-i}\right)^{\frac{\mu+|\nu|}{2}+1} e^{i\nu\gamma}\left(\frac{Q}{z}\right)^{|\nu|}\exp\left[-\frac{Q^2}{2z(z-i)}\right]M\left(\frac{|\nu|-\mu}{2}, |\nu|+1; \frac{Q^2}{2z(z-i)}\right), \quad (14)$$

$$C_{l\mu} = 2^{\frac{\mu-|\nu|}{2}}(-i)^{|\nu|+1}\frac{\Gamma\left(\frac{\mu+|\nu|}{2}+1\right)}{\Gamma(|\nu|+1)}. \quad (15)$$

where $M$ symbolizes the confluent hypergeometric function (Kummer function). The beams with complex amplitude distribution in form (14) already occurred in the theory of fork-like CGH [32,34–36,40] and were called "Kummer beams". Noteworthy, the Kummer functions of the form (14) with integer non-negative $\mu$ can be represented via sums of $\mu + 1$ terms containing modified Bessel functions ([45], pt. 7.11.1.7), which is favorable for the analysis, at least in the most practical cases of not very high $m$.

It is helpful to confront expression (15) with the diffracted beam description in the perfectly aligned case [40]. Like in Ref. [40], the complex amplitude distribution is represented by means of the Kummer beams (14) (or, equivalently, hypergeometric-Gaussian modes [37]). However, now argument of the Kummer functions $Q^2/z(z-i)$ explicitly depends on the incident beam displacement via Eqs. (13), which allows referring to beams (14) as to "displaced Kummer beams". Besides, in general case the diffracted beam is described by a superposition of the displaced Kummer beams with the same argument but with different parameters $\mu$ and $\nu$. Note that in the limit $x_s \to 0$, $y_s \to 0$ (ideal alignment), $Q \to r$, $\gamma \to \psi$, and the sum (10) contains only one term with $\mu = |m|$, $\nu = m + l$, so that result (14) coincides with formula (10) of Ref. [40]. On the other hand, if $m = 0$ (Gaussian incident beam), the superposition (10) again reduces to the single Kummer beam whose expression, in turn, appears to be equivalent to Eqs. (14) – (17) of Ref. [36]. Therefore, our results agree with more simple formulas derived before and are their immediate generalizations.

This observation partly explains the physical meaning of analytical formulas (10) – (14) but another representation of the obtained results makes it even clearer. With allowance for the definitions of $Q$ and $\gamma$ (13), one can easily derive

$$Q^2 = \left(re^{i\psi} + izr_s e^{i\phi_s}\right)\left(re^{-i\psi} + izr_s e^{-i\phi_s}\right) = r^2 - z^2 r_s^2 + 2izrr_s \cos(\psi - \phi_s) \tag{16}$$

$$e^{i\nu\gamma} Q^{|\nu|} = \left(re^{i\psi\,\mathrm{sgn}(\nu)} + izr_s e^{i\phi_s\,\mathrm{sgn}(\nu)}\right)^{|\nu|} = \left[x - x_V + i\,\mathrm{sgn}(\nu)(y - y_V)\right]^{|\nu|} \tag{17}$$

where

$$x_V = \mathrm{sgn}(\nu) z y_s, \quad y_V = -\mathrm{sgn}(\nu) z x_s. \tag{18}$$

After these transformations, from Eq. (10) we arrive at expression

$$u_l(r,\psi,z) = \frac{1}{z} \exp\left[-\frac{|\mathbf{r} - \mathbf{r}_s|^2}{2(1+iz)}\right] \sum_{\mu=0}^{|m|} \frac{\sqrt{|m|!}}{\mu!(|m|-\mu)!} \left[-r_s e^{i\phi_s\,\mathrm{sgn}(m)}\right]^{|m|-\mu}$$

$$\times C_{l\mu} \left(\frac{r}{z} e^{i\psi\,\mathrm{sgn}(\nu)} + ir_s e^{i\phi_s\,\mathrm{sgn}(\nu)}\right)^{|\nu|} M\left(\frac{|\nu|-\mu}{2}, |\nu|+1; \frac{Q^2}{2z(z-i)}\right) \tag{19}$$

which can be suitably interpreted.

First remark concerns the azimuthal dependence of the complex amplitude distribution (19). Due to Eqs. (8), (16) and (17), one can easily see that, within an inessential constant phase factor, $u_l(r,\psi,z)$ depends only on the difference $(\psi - \phi_s)$. This is a consequence of the fact that Eq. (4) is equally valid for the beam transformation performed by the spiral phase plate [47–49] and means that, in contrast to the rectangular symmetry of the "fork" CGH (see Fig. 1), all azimuthal directions of the incident beam displacements are equivalent [35,36,40]. So, when studying the displacement effects, one can accept this direction optionally (in the present work, as before [36], we choose the displacement along axis $x_a$). The exponential pre-factor describes evolution of the Gaussian envelope of the displaced incident beam; it characterizes the transverse profile of a hypothetical Gaussian beam that approaches the CGH coaxially with the real incident beam and passes the hologram without changes of its structure. Other terms can be considered as its modifications caused by diffraction and OV formation. Note that each term with $\nu \neq 0$ of the superposition in (19) possesses an isotropic OV of the order $\nu$ located at

$$x = x_V, \quad y = y_V \tag{20}$$

(see Eq. (18)) or, in the polar-coordinate representation,

$$\psi = \psi_V = \phi_s - \mathrm{sgn}(\nu)\frac{\pi}{2}, \quad r = r_V = r_s z. \tag{21}$$

According to (12), the superposition components with $\nu = 0$ can exist only if simultaneously

$$ml < 0, \quad |m| \geq |l|. \tag{22}$$

While condition (22) violates, the whole diffracted beam (19) possesses an OV located at the point specified by Eqs. (20) and (18) or Eq. (21). Its order is determined by the power term in parentheses of Eq. (19) with the lowest non-zero power $|v|$ and equals to

$$v_V = \begin{cases} l, & \text{if } ml > 0, \\ l+m, & \text{if } ml < 0. \end{cases}$$

For example, such OVs (corresponding to $v_V = 1$ and $-1$) are shown in 3$^{rd}$ and 4$^{th}$ rows of Fig. 2 by asterisks. However, there can exist other OVs in the beam (they are also seen in the 3$^{rd}$ and 4$^{th}$ rows of Fig. 2), stipulated by the interference between different components of superposition (19), whose positions cannot be found by general analysis. Likewise, under conditions (22), the term corresponding to $\mu = |m|$ ($v = 0$) with no OV at (20) is present, and formula (19) provides no analytical prediction as to positions and other parameters of the diffracted beam OVs. In this case, all zeros of the beam amplitude appear as a result of destructive interference of the separate Kummer beams entering the superposition (19).

The exponential pre-factor of (19) decays exponentially at the beam periphery ($r \to \infty$) which however can be misleading in consideration of the whole amplitude behavior. In fact, the pre-factor falloff can be compensated by growth of the Kummer functions. The resulting effect can be seen from the asymptotic representation of function (19)

$$u_l(r,\psi,z)\Big|_{r\to\infty} = \sum_{\mu=0}^{|m|} \frac{\sqrt{|m|!}}{\mu!(|m|-\mu)!} \left[-r_s e^{i\phi_s \operatorname{sgn}(m)}\right]^{|m|-\mu} e^{iv\psi} \qquad (23)$$

$$\times \left\{ (-i)^{|v|+1} \frac{\Gamma\left(\frac{\mu+|v|}{2}+1\right)}{\Gamma\left(\frac{|v|-\mu}{2}\right)} \exp\left(i\frac{r^2}{2z} - \frac{r_s^2}{2}\right) \frac{(2z)^{\mu+1}}{r^{\mu+2}} \left[1 + O(r^{-2})\right] \qquad (24) \right.$$

$$\left. + \frac{r^\mu}{(1+i\zeta)^{\mu+1}} \exp\left[-\frac{|\mathbf{r}-\mathbf{r}_s|^2}{2(1+i\zeta)}\right]\left[1 + O(r^{-2})\right] \right\}. \qquad (25)$$

which also is a generalization of corresponding result for "ordinary" Kummer beams given by Eqs. (13) and (14) of Ref. [40] (by the way, Eq. (14) of [40] contained an inessential mistake that is corrected in Eq. (25) above). Each member of the sum is asymptotically represented by two summands. The first one (Eq. (24)) corresponds to the diverging spherical wave that originates from the bifurcation point of the grating (see Fig. 1); the second one (Eq. (25)) expresses the "regular" vortex wave propagation. Contribution (24) usually dominates unless it does not vanish, i.e. unless condition $(|v|-\mu)/2 = 0, -1, -2, ...$, or

$$\mu - |l + \mu \operatorname{sgn}(m)| = 0, 2, 4, ... \tag{26}$$

does not hold. In this situation, at $r \to \infty$ the beam amplitude falls down in the power-law way, much slower than the exponential decay of the incident LG beam. Condition (26) is always fulfilled (and the edge wave (24) does not emerge) if $l = 0$, that is, in the zero-order diffracted beam when no OV transformation takes place. On the contrary, at $l \neq 0$ and $r_s \neq 0$ the sum (23) or (10) inevitably contains a term with $\mu = 0$ for which condition (26) violates. Therefore, displacement of the incident LG beam with respect to the hologram center, among other consequences, inevitably causes appearance of the edge wave with all the accompanying effects [36,40]. The exponential transverse confinement of the incident beam is replaced by the less efficient power-law confinement of the diffracted beam, interference between the regular and the edge waves produces the characteristic "ripple" pattern in both the intensity and phase distributions, etc. This is not surprising because requirements for the edge wave (24) to be suppressed include, first of all, zero intensity of the incident beam at the bifurcation point [40], which is not the case for displaced LG beams.

In the far field ($z \to \infty$), the result (19) can be expressed via the angular variable

$$\mathbf{p} = \frac{\mathbf{r}}{z} \tag{27}$$

which determines the angle of propagation in units of the 'prototype' divergence angle $(kb)^{-1}$:

$$u_l(r,\psi,z) \to U_l(p,\psi,z) = \frac{1}{z} \exp\left(\frac{iz-1}{2} p^2\right) \sum_{\mu=0}^{|m|} \frac{\sqrt{|m|!}}{\mu!(|m|-\mu)!} \left[-r_s e^{i\phi_s \operatorname{sgn}(m)}\right]^{|m|-\mu}$$

$$\times C_{l\mu} \left(pe^{i\psi \operatorname{sgn}(\nu)} + ir_s e^{i\phi_s \operatorname{sgn}(\nu)}\right)^{|\nu|} M\left(\frac{|\nu|-\mu}{2}, |\nu|+1; \frac{1}{2}\left(pe^{i\psi} + ir_s e^{i\phi_s}\right)\left(pe^{-i\psi} + ir_s e^{-i\phi_s}\right)\right) \tag{28}$$

where

$$p = |\mathbf{p}|.$$

Like Eq. (19), the far-field distribution (28) depends on $\psi$ only via the difference $(\psi - \phi_s)$; additionally, replacement $\psi \to \pi + 2\phi_s - \psi$ transforms (28) to the complex conjugate. Consequently, the far-field intensity pattern possesses the mirror symmetry with respect to axis $\psi = \phi_s \pm \pi/2$ while the phase distribution changes sign. Illustrations of such a behavior (at $\phi_s = \pi$) are provided below in the last columns of Figs. 2, 3 and in Fig. 8.

### 4. Numerical examples and calculations

General view of the diffracted beam profiles for some combinations of the incident OV order $m$ and the embedded topological charge of the CGS $l$ is given in Figs. 2 and 3; the incident beam is supposed to be displaced to the negative axis $x_a$ ($\phi_s = \pi$). All figures represent the patterns seen against the beam propagation. The 1$^{st}$ and 2$^{nd}$ rows of Fig. 2 (Figs 2 (aa)–(ad) and (ba)–(bd)) describe the intensity distribution of beams with "healed" singularity ($m = -l$) [40,47] while the 3$^{rd}$ and 4$^{th}$ rows (Figs 2 (ca)–(cd) and (da)–(dd)) show behavior of the beams where the singularity of the incident beam is transformed but survive. Typical phase profiles can be seen in Fig. 3 via the constant phase contours; for simplicity, only one situation with "healed" singularity and one with "surviving" OV are illustrated, and the far-field pictures (Figs. 3c, f) are "purified" from the overall wavefront curvature factor $\exp(izp^2/2)$ (see Eq. (28)).

At early stages of evolution (first column of Fig. 2), the diffracted beam intensity patterns are rather similar and reproduce the incident LG$_{01}$ pattern distorted by the OV generated with the CGH and modulated by the ripple structure appearing due to interference between the "regular" and "diverging" waves corresponding to terms (25) and (24) of the complex amplitude asymptotic expression [35,40,47]. The ripples are as well seen in the phase distributions given by the first column of Fig. 3. The "proper" OV of the incident beam is situated near the point $x = x_s$, $y = y_s$ ($x = -1$, $y = 0$ in Figs. 1, 2) and is distinctly separated from the OV "imparted" by the CGH (near the frame origin). The symmetry breakdown associated with the incident beam displacement causes that high-order imparted OVs (as well as high-order "proper" OVs) decompose into sets of single-charged ones (this is not perceptible in the intensity patterns of Fig. 2(ba), (da) but is well seen in Figs. 3–5 displaying the dynamics of the OVs within the diffracted beam cross sections).

With further evolution, the ripple structure expectedly [35,36,40,47–49] becomes less articulate (2$^{nd}$ column of Figs. 2, 3) and disappears in the far field (last column of Figs. 2, 3). Simultaneously, the beam transverse pattern evolves to the stable far-field form. In this process, the diffracted beams with surviving OV show features of the transverse profile rotation in direction determined by the overall transverse energy circulation (positive – counter-clockwise – in 3$^{rd}$ row of Fig. 2 and negative in the 4$^{th}$ row), which is characteristic for OV beams with broken symmetry [5,36,39]. In the far field, the beams with "healed" singularity (panels (ad), (bd) of Fig. 3) show better energy concentration in the main lobe even in comparison with the hypothetic case as if an incident Gaussian beam "freely" propagates through the CGH into the chosen diffraction order without accepting the phase singularity (see light rings in the far-field column of Fig. 2). However, in contrast to the perfect-alignment case of Ref. [40], now the additional intensity maxima appear (bright spots above the main lobes in panels (ad) and (bd)) whose role grows with growing $r_s$ and $|m| = |l|$.

The most impressive differences from the known pictures of the diffracted beam evolution in the perfectly aligned case [40] or when the incident beam is Gaussian [36] are associated with the behavior of individual OVs within the beam cross section. In the "singularity healing" situations (1st and 2nd rows of Fig. 2 and 1st row of Fig. 3), lack of symmetry caused by misalignment prevents from perfect disappearance of the singularity. The "proper" and "imparted" singularities are spatially separated and usually cancel each other only "on average" for the beam as a whole. Sometimes they "temporarily" annihilate, as OVs $V_1$ and $V_2$ in Fig. 4a and Fig. 5b or $V_4$ and $V_5$ in Fig. 4d, but typically they migrate from the beam center (like $V_1$ in Figs. 4b–d) where the intensity lobe is formed instead. During the whole evolution, some OVs are continuously present somewhere near the intensity lobes within the so called "active" beam area[1] (see Figs. 3c and 4). Besides, plural strongly anisotropic OVs appear at the beam periphery in connection with the ripple structure, as is seen in Fig. 4b, e.

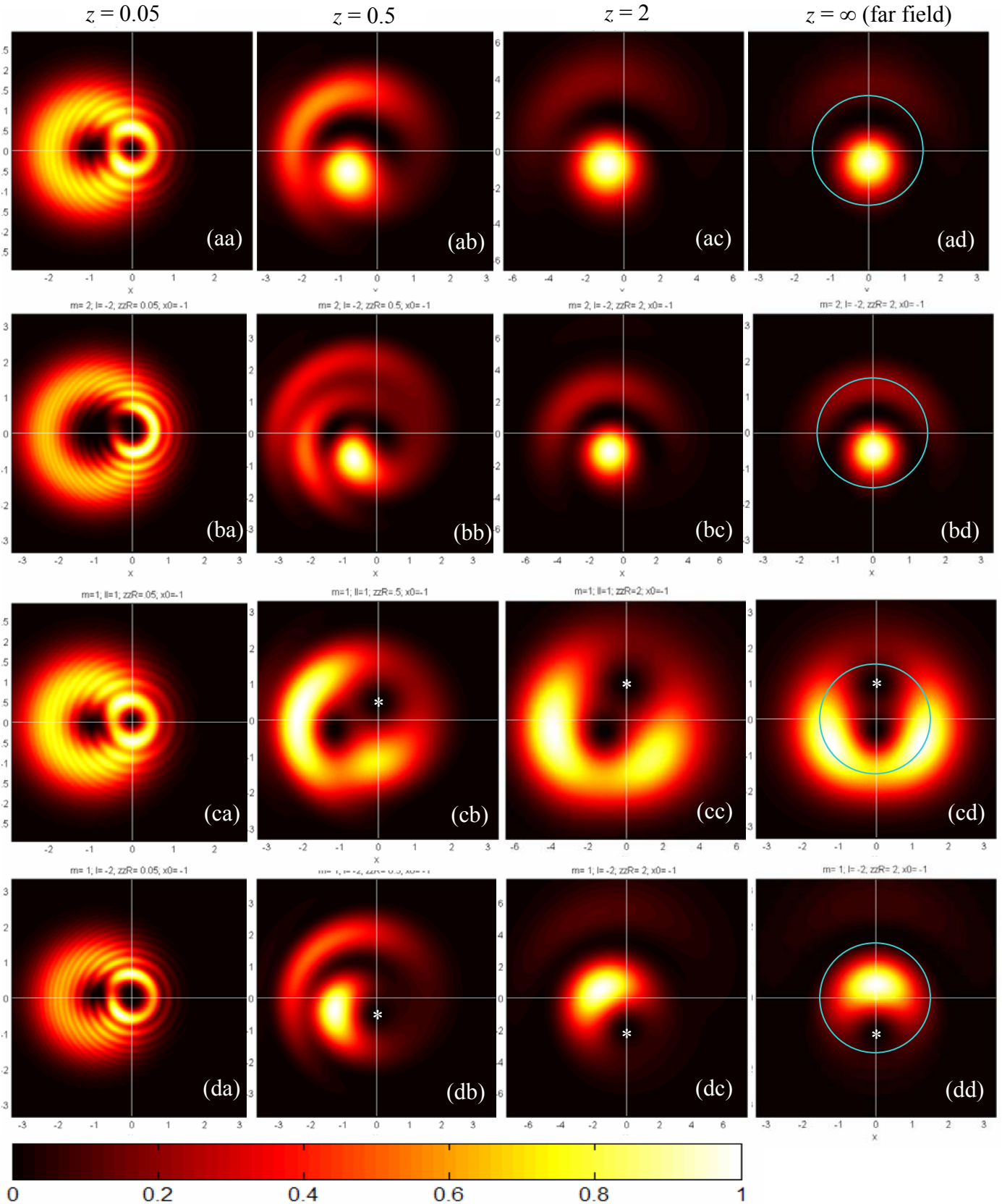

Fig. 2. Transverse intensity profiles of the diffracted beams obtained from the CGH when the incident beam displacement components equal to $x_s = -1$, $y_s = 0$. Each column corresponds to certain propagation distance marked above the column, parameter values: (1st row) $m = 1$, $l = -1$; (2nd row) $m = 2$, $l = -2$; (3rd row) $m = 1$, $l = 1$; (4th row) $m = 1$, $l = -2$. Asterisks denote OVs situated in points specified by (18), (20); in the last column, for comparison, 10%-maximum equal intensity contours are shown for a hypothetic Gaussian beam which arrive at the CGS coaxially with the incident beam and experiences no structure transformation when crossing the CGH.

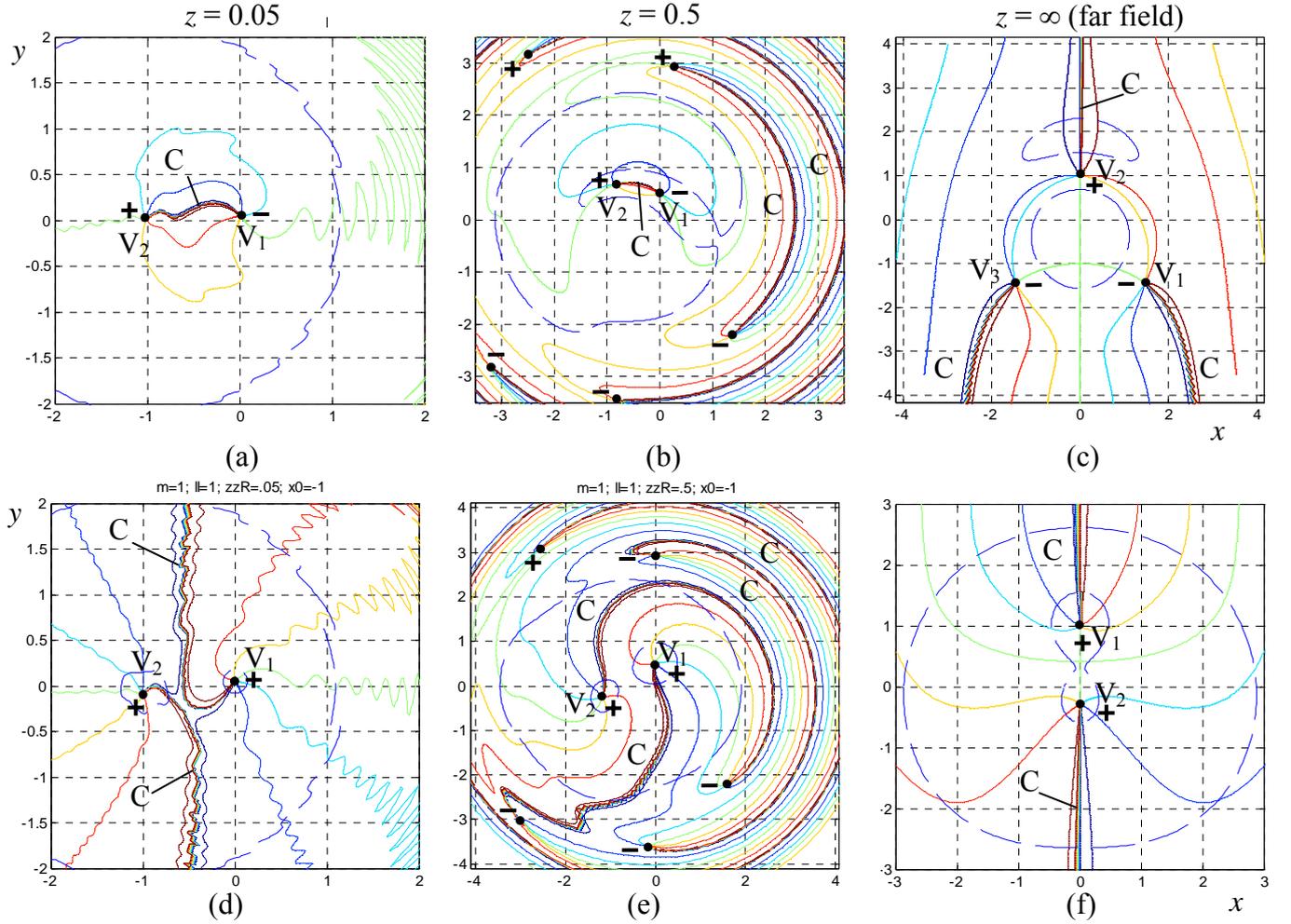

Fig. 3. Contours of constant phase in transverse sections of the diffracted beams obtained from the CGH when the incident beam displacement components equal to $x_s = -1$, $y_s = 0$. Columns correspond to 1st, 2nd and 4th columns of Fig. 2 (propagation distance is indicated above), rows correspond to the 1st ($m = 1$, $l = -1$) and 3rd ($m = 1$, $l = 1$) rows of Fig. 2. At the OV positions (●) different equiphase contours converge; cuts (C) of the phase surface are seen by visual merging of several phase contours. Signs "+" and "–" denote positive (counter-clockwise energy circulation) and negative OVs; $V_1$, $V_2$ and $V_3$ are the vortices discussed in the text (see also Figs. 4, 5). Dashed lines: contours of the current beam intensity at the level 0.1 of maximum.

With further propagation, most of them move to the transverse infinity but some OVs migrate to the active area and take part in interesting topological reactions (in total, they constitute what can be called "OV skeleton" [6] of the beam). This sort of evolution is displayed in Figs. 4 and 5 where current positions of the OVs are presented in normalized coordinates

$$\frac{x}{\sqrt{1+z^2}}, \ \frac{y}{\sqrt{1+z^2}}. \tag{29}$$

The OV dynamics essentially depends on the initial beam displacement (Figs. 4a–c). At relatively small $r_s$ (in conditions of Fig. 4a–c, $r_s < 1$), the "proper" and "imparted" OVs $V_2$ and $V_1$ start to move bowingly, then converge and annihilate at $z = 0.29$ (point A in Fig. 4a). For some propagation distance there is no OV in the active beam area, but further (in Fig. 4a, at $z = 2.8$) a pair of OVs are born in point B. These can be considered as reviving $V_1$ and $V_2$. Meanwhile, at $z = 3.0$ a new OV $V_3$ enters the active area from the beam periphery. Note that dependence of the OVs' behavior on the incident beam shift $r_s$ is rather steep: already at $r_s = 1.2$ the evolution if the OV system looks quite differently, OVs neither are born nor annihilate, and vortex $V_3$ appears noticeably earlier (at $z = 1$ in Fig. 4c). The case of $r_s = 1.0$ forms a boundary between the two evolution types: vortices $V_1$ and $V_2$ visually run into each other at $z = 1$ (Fig. 4b) but immediately afterwards two OVs diverge and it is equally lawful to consider this process as evolution of the same vortices $V_1$ and $V_2$ or as birth of a new pair. Despite the different ways of evolution, the "final" picture of the far-field OV skeleton is qualitatively the same and corresponds to the phase picture presented in Fig. 3c.

The case of $m = -l = 1$ represents a rather simple evolution pattern. In essence, the similar processes of the OVs' evolution occur in situations with higher $|m|$ and $|l|$ but the picture becomes more complicated and rich of details. If $m = -l = 2$ (Fig. 4d), already at $z = 0.05$ the two-charged proper and imparted OVs are split in pairs of closely neighboring single-charged ones. Then, positive and negative OVs $V_4$ and $V_5$ approach together and annihilate at $z = 0.15$ while their counterparts $V_1$ and $V_2$ survive during the whole beam evolution; an additional vortex $V_3$ arrives at the beam active area from the periphery. The triple of vortices $V_1$, $V_2$ and $V_3$ behave quite analogously to the OVs of the same names in Fig. 4a–c. There are many other OVs but, as a rule, they are situated at the far beam periphery where the intensity is negligible. Possibly, only one of them can play some role in the beam characterization – it is $V_6$ that migrate over the whole beam cross section and can approach closely to the noticeable-intensity areas in the course of evolution.

---

[1] We apply this term to the parts of the beam cross section, in or around which the radiation intensity is high enough to observations. Of course, such a definition is not strict and depends on specific experimental conditions; by its introducing we emphasize that only few OVs lying not very far from the beam bright spots will be taken into account.

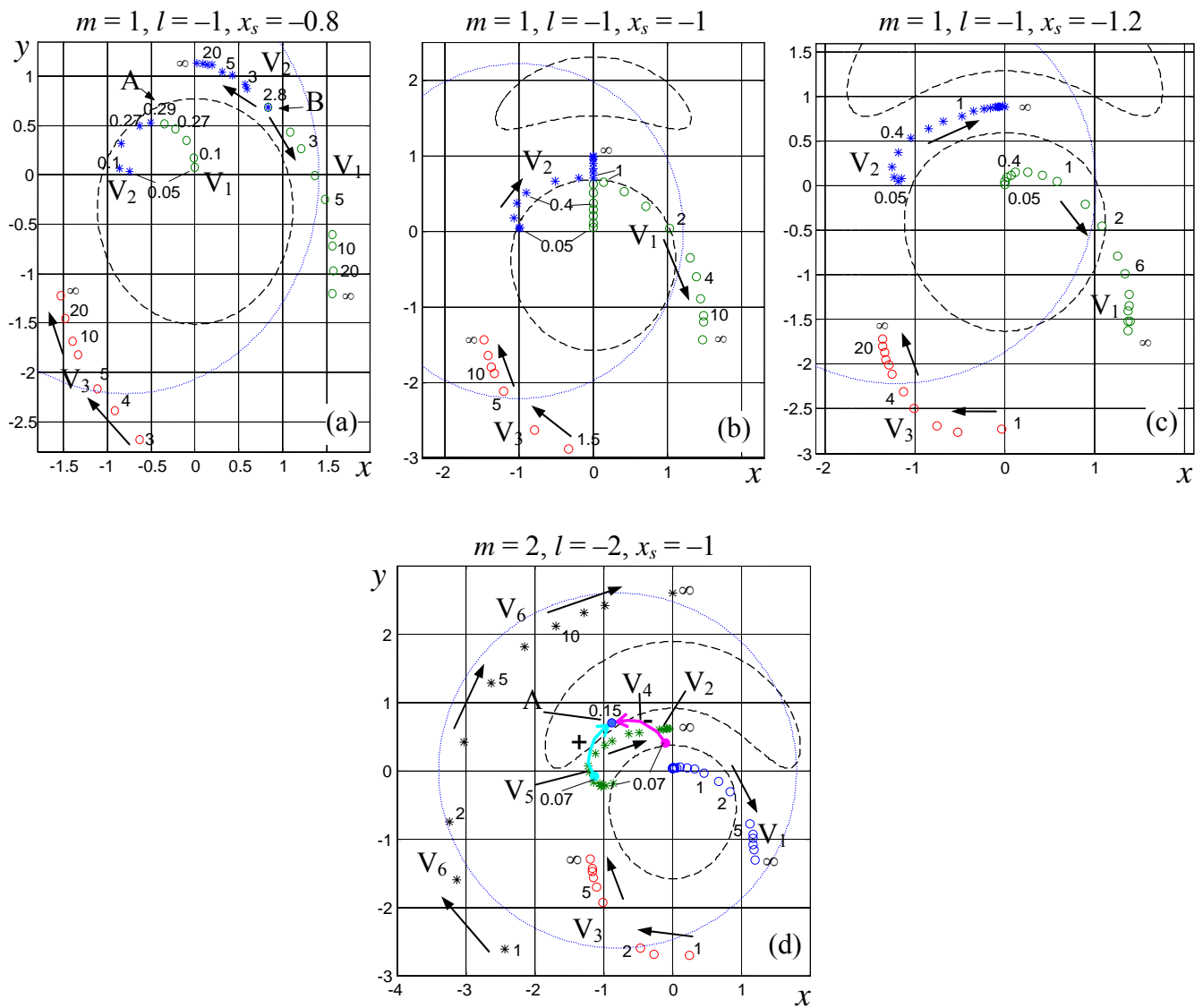

Fig. 4. Maps of the OVs' evolution in normalized coordinates (29) for diffracted beams with $m = 1$, $l = -1$ (a–c) and $m = 2$, $l = -2$ (d); values of $x_s$ are different and indicated above the panels. Current positions of the OV cores are denoted by asterisks (positive OVs) and circles (negative OVs); figures near points denote the propagation distances $z$, arrows show directions of OV motion during the beam propagation; points of annihilation (A) and birth (B) are shown. Trajectories of vortices $V_4$ and $V_5$ in (d) are shown by solid lines, their signs are marked by "+" and "−". Dashed (dotted) lines: contours of the far-field (incident beam) intensity at the level 0.1 of the maximum.

The OV distribution for the beams with "surviving" singularity (Fig. 5 describes the beams presented in 3$^{rd}$ and 4$^{th}$ rows of Fig. 2 and 2$^{nd}$ row of Fig. 3) confirm the conclusion that richness and diversity of the OV skeleton quickly grows with the growing $|m|$ and $|l|$. In case $m = l = 1$ the phase pattern (Fig. 3d–f and Fig. 5a) is perhaps the simplest, at least for beams considered in this paper. The active area of the beam transverse profile contains only two positive OVs which move almost straightly from initial positions, dictated by the incident beam displacement with respect to the CGH center, to the far-field locations. Small irregularities in the initial region of trajectory of OV $V_2$ in Fig. 5a can be ascribed to the influence of the ripple structure of Fig. 2 and 3 (1$^{st}$ column); in fact, similar irregularities can be seen in other cases but in the considered figure they are especially well expressed. The "imparted" OV $V_1$ moves along the $y$-directed straight line in accord with Eq. (18), (20) and (21) for $x_s = -1$, $y_s = 0$, the same reason stipulates the similar behavior of $V_1$ in Fig. 5b. However, other features of Fig. 5b are quite different. In the corresponding case $m = 1$, $l = -2$, at the earliest stages of evolution, three OVs exist (the proper one $V_2$ and two "fragments" of the initially two-charged OV imparted by the hologram). Besides, a lot of anisotropic OVs appear outside the beam active area due to interference with the divergent wave, similar to OVs shown in Fig. 4b, e. In the course of further propagation, $V_1$ and $V_2$ annihilate at $z = 0.3$ but OVs $V_4$, $V_7$ and $V_8$ approach the active area from the beam periphery. Then $V_7$ and $V_8$ move rather quickly around the beam cross section and annihilate at $z = 0.8$ while $V_4$ remains during the whole beam "life story". Afterwards, for a long evolution period only vortices $V_3$ and $V_4$ are present in the active area and they slowly move to their far-field positions. Surprisingly, only at $z = 20$, a new OV pair $V_5$ and $V_6$ is born. Their further migration leads to formation of the final far-field picture with $V_5$ and $V_3$ situated at the $y$ axis and $V_4$ and $V_6$ placed symmetrically on each side (points marked $\infty$ in Fig. 5b).

Note that all far-field distributions of the OVs of Figs. 3c, e, 4 and 5 appear to be symmetrical with respect to the $y$ axis, as it follows from the analytical expression (28). Another important observation connected to Figs. 3–5 is that essential qualitative changes of the phase profiles can occur in "very far-field" conditions when the intensity distribution practically reaches its far-field form and no noticeable variations of the intensity profile can be expected. This is demonstrated by birth of the OV pair in point B in Fig. 4a, migration of OVs $V_3$ in Fig. 4a–c and $V_6$ in Fig. 4d, and, especially spectacular – by birth of $V_5$ and $V_6$ at $z = 20$ in Fig. 5b.

Positions of OVs are examples of so called "featuring points" of the beam transverse structure – special points of the intensity and phase distributions that are (relatively) stable and provide general ideas on the beam spatial shape and location [36]. Other such points are positions of the intensity maxima $(x_m, y_m)$ and coordinates of the "centroid", or "center of gravity" (CG) defined as

$$\begin{Bmatrix} x_0(z) \\ y_0(z) \end{Bmatrix} = \frac{\int \begin{Bmatrix} x \\ y \end{Bmatrix} |u_l(x,y,z)|^2 \, dx\, dy}{\int |u_l(x,y,z)|^2 \, dx\, dy} = \frac{\int r \begin{Bmatrix} \cos\psi \\ \sin\psi \end{Bmatrix} |u_l(r,\psi,z)|^2 \, r\, dr\, d\psi}{\int |u_l(r,\psi,z)|^2 \, r\, dr\, d\psi}. \qquad (30)$$

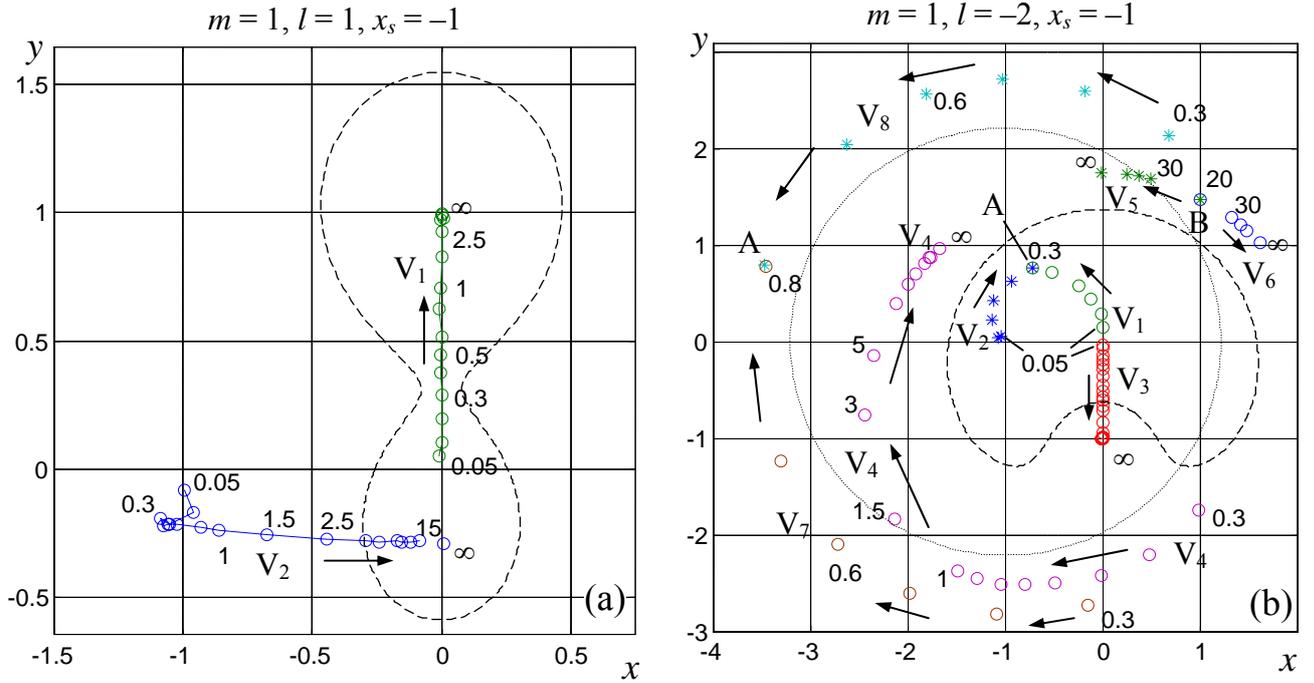

Fig. 5. The same as Fig. 4 but for diffracted beams with $m = 1$, $l = 1$ (a) and $m = 1$, $l = –2$ (b); see further explanations in the text and in caption to Fig. 4.

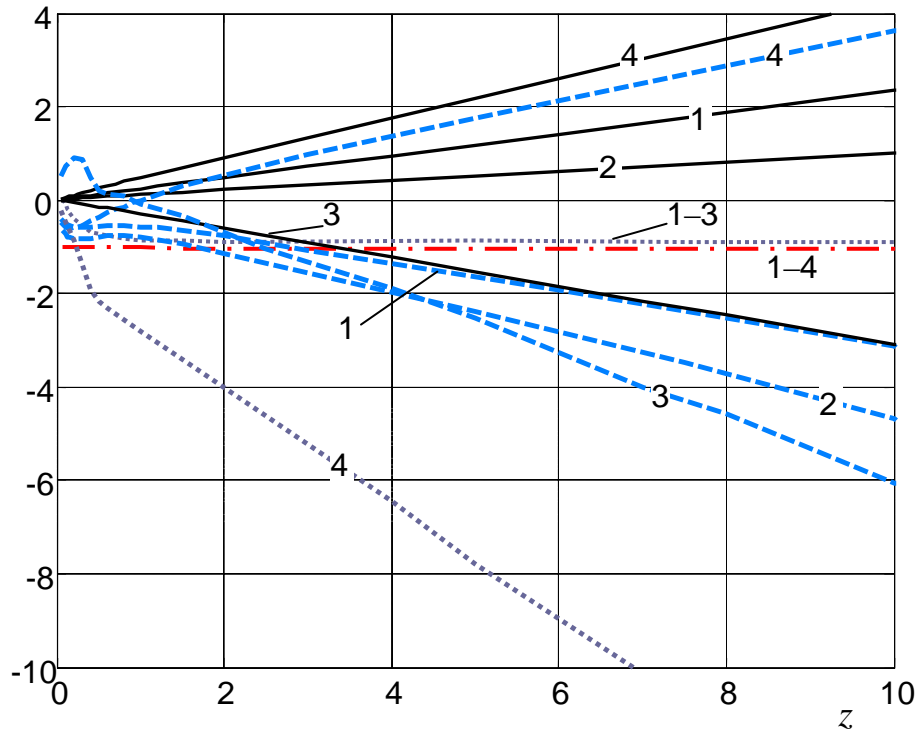

Fig. 6. Off-axial displacements of the featuring points of the diffracted beam transverse intensity profile vs propagation distance: (1) $m = 1$, $l = –1$; (2) $m = 2$, $l = –2$; (3) $m = 1$, $l = 1$; (4) $m = 1$, $l = –2$. Solid lines: $y_0$ (30), dashed lines: $y_m$, dotted lines: $x_m$, dash-dot lines: $x_0$; dotted lines $1 – 3$ visually coincide as well as dash-dot lines $1 – 4$.

Corresponding information is given in Fig. 6. In agreement with earlier results obtained for the Gaussian incident beam, all the dependences are practically linear. This is quite natural for the CG components since the CG propagates rectilinearly in free space. The fact that only the CG component $y_0$, orthogonal to the incident beam displacement, grows upon propagation (in conditions of Fig. 6, $x_0$ always preserves its initial value obviously coinciding with the incident beam displacement $x_s$), has its prototype in the behavior of the OV beams generated by displaced Gaussian beams and can be attributed to the same reason connected to the special feature of the helical wavefront deformation imposed by the CGH [36].

As for coordinates of intensity maximum $x_m$ and $y_m$, their linear dependence on $z$ seems to be approximate. However, its visible violations are observed only at small $z$, which can be attributed to the influence of the "ripple" structure. Linear trajectories of the maximum points seem to have no simple geometric interpretation.

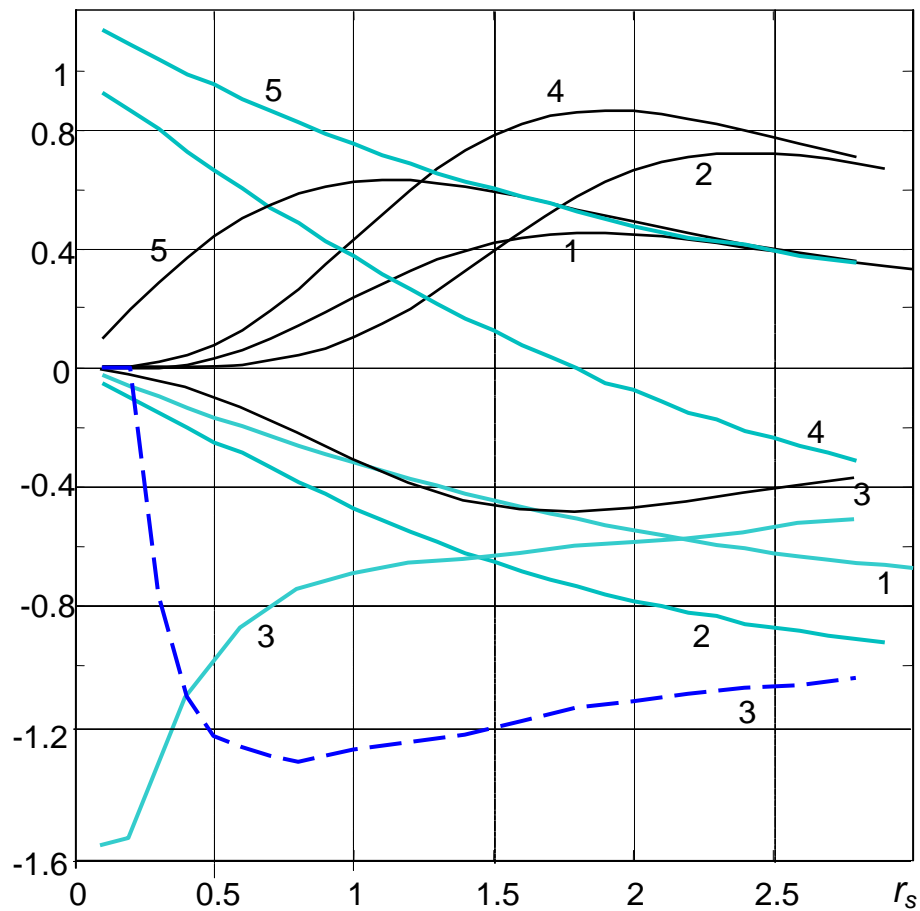

Fig. 7. Far-field positions of featuring points of the diffracted beam transverse intensity profiles vs the incident beam misalignment: (1) $m = 1$, $l = -1$; (2) $m = 2$, $l = -2$; (3) $m = 1$, $l = 1$; (4) $m = 1$, $l = -2$; (5) $m = 0$, $l = 1$. Black lines: $y_0$ (30), light lines: $y_m$, dashed line: coordinate $x_m$ of the left bright maximum of the profile of Fig. 2(cd).

Due to linear character of dependences demonstrated by Fig. 6, current positions of any featuring point can be exhaustively described by the slope of the corresponding line, which coincides with the featuring point deviation in the far field expressed via angular variable (27). This slope depends on the incident beam shift and knowledge of this dependence is useful for analysis of the diffracted beam sensitivity to deviations from the ideal scheme of the OV beam transformation in the CGH. In Fig. 7, the result obtained for beams whose behavior was discussed in the previous figures, are compared to the misalignment sensitivity of the diffracted OV beam generated by the CGH from the incident Gaussian beam with $m = 0$, $l = 1$. Here, again, the components orthogonal to the incident beam shift are of main interest: for all far-field profiles of Fig. 2 (last column), $x$-coordinates of the featuring points are zeros. One can show that, generally, the transformed Gaussian beam is more sensitive to small misalignments (slope of curve 5 is much higher than that of all other black curves). The case of $m = 1$, $l = 1$ is an exception: here $x_m = 0$ only at very small $r_s$ when the initial LG-ring of the incident beam is almost non-deformed; if $r_s > 0.2$, two symmetric maximums appear in both sides of the ring (a well developed deformation picture is seen in Fig. 2(cd)). In the transient region $0.2 < r_s < 0.5$ the slope of the dashed curve $dx_m/dr_s$ is rather high so the far-field position of the intensity maximums of the diffracted beam with $m = 1$, $l = 1$ can serve a sensitive indicator of small displacements of the incident beam or/and the CGH.

Of course, phase profiles of the diffracted beams may experience essential reconstruction with changing $r_s$. For the far-field cross sections it is demonstrated by Fig. 8 where the distribution and migration of the individual OVs over the beam cross section is presented. At $r_s = 0$, there are no OVs in the active area but a ring of zero intensity (similar to the usual Airy ring) with radius $p \approx 1.75$ ($p \approx 1.4$) for conditions of Fig. 8a (8b) exists [40]. Upon the smallest displacement, four OVs are formed instead, lying symmetrically on this ring; three of them can be associated with OVs $V_1$, $V_2$ and $V_3$ introduced in Fig. 4, the 4$^{th}$ one is denoted in Figs. 8a, b as $V_7$ (we try to preserve the nomenclature of OVs in which notations $V_4 - V_6$ are already assigned in Fig. 4). Then, with growing $r_s$, these vortices move towards negative $y$ direction, $V_1$ and $V_3$ approaching closer to the $y$ axis. OVs $V_1$, $V_2$ and $V_3$ survive up to rather large misalignments while positive vortex $V_7$ collide with the negative one $V_8$, that enters the active area from the periphery, and they annihilate (point A in Figs. 8a, b). By this, qualitative evolution of the "OV skeleton" for a beam produced by the CGH with $m = -l = 1$ ends. In the beam with $m = -l = 2$ the 'life story' looks more complicated (Fig. 8b). Already at $r_s = 0.2$, negative OV $V_6$ enters on the scene and moves downwards; at $r_s = 1.19$ (point R in Fig. 8b) it becomes extremely anisotropic, "emits" a pair of negative OVs $V_{10}$ and $V_{11}$ and converts into positive vortex $V_7$. All these OVs survive during the whole range of $r_s$ variation.

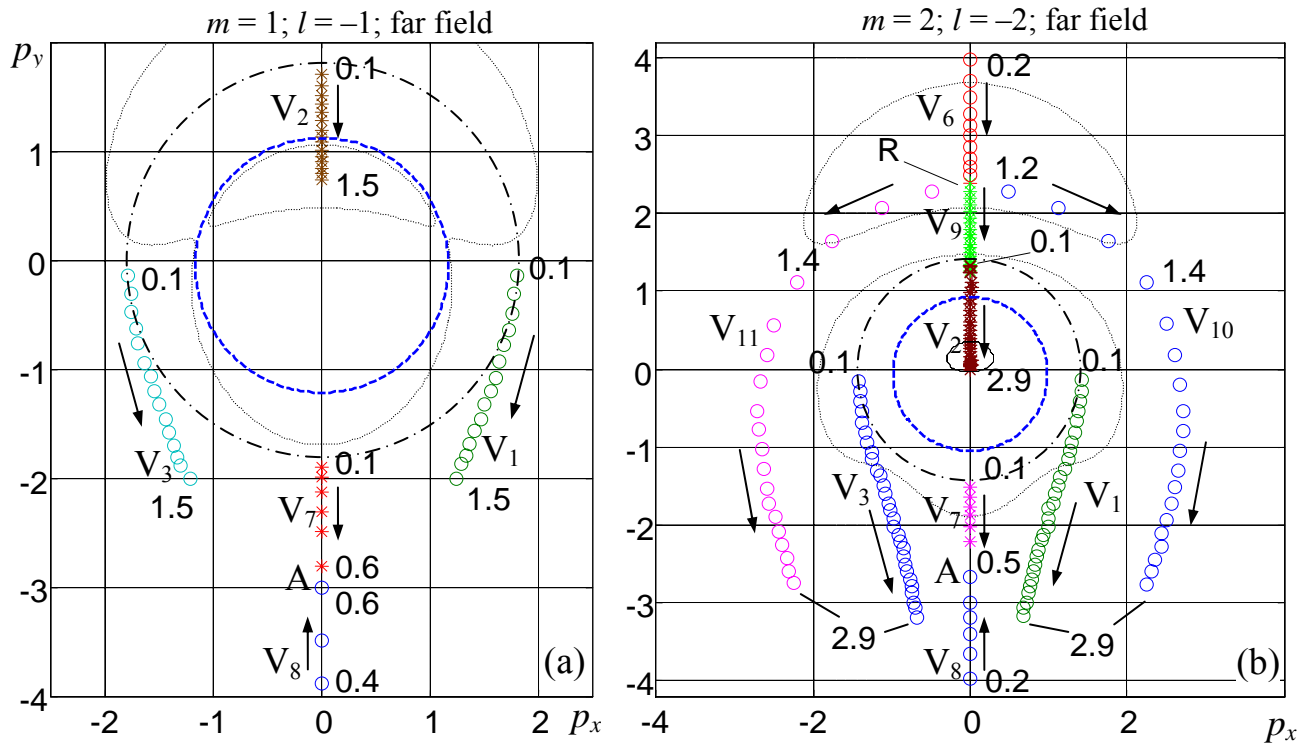

Fig. 8. Far-field maps of the OVs' distribution for different incident beam displacements $r_s$ ($\phi_s = \pi$) for diffracted beams with $m = 1$, $l = -1$ (a) and $m = 2$, $l = -2$ (b). Current positions of the OV cores are denoted by asterisks (positive OVs) and circles (negative OVs); figures near points denote the corresponding values of $r_s$, arrows show directions of OV motion with growing $r_s$. Dashed lines: contours of the 10% of maximum intensity level for $r_s = 0.1$; dotted lines: the same for $r_s = 1.5$ (panel (a)) and $r_s = 2.0$ (panel (b)); dash-dot lines: first dark ring for $r_s = 0$. Points of topological reaction are indicated: (A) annihilation of $V_7$ and $V_8$, (R) transformation $V_6 \to V_9 + V_{10} + V_{11}$.

### 4. Conclusion

Results of this paper can be divided in two main groups. The first group consists of analytical theory starting with Eq. (4) and until Eq. (28), which provides reliable and suitable means for investigation of diverse situations occurring in transformations of LG$_{0m}$ beams by the "fork" holograms. Moreover, the whole theory is equally valid for description of the beam transformations performed by spiral phase plates [47,49] with integer phase step. However, cases of CGH or spiral plates with fractional embedded topological charge [25,32,33] need, generally, more exquisite analytical means (see, e.g., Refs. [33,47]). In the present form, the theory is directly applicable to somewhat simplified situations where the incident beam axis is normal to the CGH and the CGH input plane coincides with the waist of the incident beam. The first limitation was briefly discussed in Sec. 2 and can be easily overcome by adding the supplementary phase factor and introducing standard transformations to the argument of the complex amplitude distributions (19) or (28) [36]. The second one is equally inessential and case when the incident LG beam possesses arbitrary wavefront curvature can be exhaustively described by the presented formulae after corresponding scaling and translation of the spatial arguments [36,39].

Other results, presented in Sec. 3, are, in fact, examples of the theory application. However, despite a very limited number of considered concrete situations, some observations can be useful for various practical problems of the OV generation and usage. In particular, known dependencies of the diffracted beam structure on the incident beam misalignments are helpful for assigning tolerances and required accuracy of optical systems designed for sensing and analysis of OV beams [19–25]. On the other hand, changes of the diffracted beam profile due to misalignments can be utilized for exact measurement of small geometrical displacements, deformations or disagreements [10–15]. To this purpose, the situations where small changes of $r_s$ lead to substantial qualitative reconstruction of the output beam profile (see, for example, the OV skeleton transformations in Figs. 4a–c, behavior of the dashed curve in Fig. 7 at $0.2 < r_s < 0.5$, points of topological reactions in Fig. 8, etc.) look quite promising. Further search of similar conditions favorable to measuring one or another parameter can be performed on the presented theoretical basis.

And, finally, the results of this paper can be employed for generation of OV beams with some special configurations and with interesting and instructive behavior of the OV skeleton, topological reactions, etc.